\RequirePackage{fix-cm}
\documentclass[twocolumn,epjc3]{svjour3}
\smartqed
\RequirePackage{graphicx}
\RequirePackage{amssymb}
\RequirePackage{mathtools}
\RequirePackage{color}
\RequirePackage{siunitx}
\RequirePackage{orcidlink}
\RequirePackage[utf8]{inputenc}
\RequirePackage{microtype}

\journalname{Eur. Phys. J. C}

\begin{document}

\title{Detector Performance and Dark Matter Sensitivity of CYGNO
with HFO-1234ze Gas Admixtures}

\author{
  F.D. Amaro\thanksref{addr1}\orcidlink{0000-0001-7315-0550} \and
  R. Antonietti\thanksref{addr2,addr3}\orcidlink{0009-0009-2568-8247} \and
  E. Baracchini\thanksref{addr4,addr5}\orcidlink{0000-0003-4686-128X} \and
  L. Benussi\thanksref{addr6}\orcidlink{0000-0002-2363-8889} \and
  F.M. Brunbauer\thanksref{addr14}\orcidlink{0000-0003-0514-8189} \and
  C. Capoccia\thanksref{addr6}\orcidlink{0009-0008-5919-3130} \and
  M. Caponero\thanksref{addr6,addr7}\orcidlink{0000-0002-5728-3123} \and
  L.G.M de Carvalho\thanksref{addr8}\orcidlink{0009-0003-5836-4771} \and
  G. Cavoto\thanksref{addr9,addr10}\orcidlink{0000-0003-2161-918X} \and
  I.A. Costa\thanksref{addr6}\orcidlink{0000-0002-3064-8305} \and
  A. Croce\thanksref{addr6} \and
  M. D'Astolfo\thanksref{addr4,addr5,e1}\orcidlink{0009-0000-9817-6693} \and
  G. D'Imperio\thanksref{addr10}\orcidlink{0000-0002-2945-0983} \and
  E. Dan\`e\thanksref{addr6}\orcidlink{0000-0002-7220-9984} \and
  G. Dho\thanksref{addr6,e2}\orcidlink{0000-0001-9454-9894} \and
  E. Di Marco\thanksref{addr10}\orcidlink{0000-0002-5920-2438} \and
  J.M.F. dos Santos\thanksref{addr1}\orcidlink{0000-0002-8841-6523} \and
  D. Fiorina\thanksref{addr4,addr5,e3}\orcidlink{0000-0002-7104-257X} \and
  F. Iacoangeli\thanksref{addr10}\orcidlink{0000-0003-0808-585} \and
  Z. Islam\thanksref{addr4,addr5}\orcidlink{0000-0003-4611-839X} \and
  E. Kemp\thanksref{addr11}\orcidlink{0000-0001-5311-1300} \and
  H.P. Lima Jr\thanksref{addr4,addr5}\orcidlink{0000-0001-7398-3237} \and
  G. Maccarrone\thanksref{addr6}\orcidlink{0000-0002-7234-9522} \and
  R.D.P. Mano\thanksref{addr1}\orcidlink{0000-0003-2920-7067} \and
  D.J.G. Marques\thanksref{addr4,addr5}\orcidlink{0000-0002-0013-6341} \and
  G. Mazzitelli\thanksref{addr6}\orcidlink{0000-0003-2830-4359} \and
  P. Meloni\thanksref{addr2,addr3}\orcidlink{0009-0001-7634-370X} \and
  A. Messina\thanksref{addr9,addr10}\orcidlink{0000-0003-1195-6780} \and
  C.M.B. Monteiro\thanksref{addr1}\orcidlink{0000-0002-1912-2804} \and
  R.A. Nobrega\thanksref{addr8}\orcidlink{0000-0001-5199-308X} \and
  E. Olivieri\thanksref{addr14}\orcidlink{0000-0002-0832-6975} \and
  I.F. Pains\thanksref{addr8}\orcidlink{0009-0004-0851-6308} \and
  E. Paoletti\thanksref{addr6} \and
  F. Petrucci\thanksref{addr2,addr3}\orcidlink{0000-0002-5278-2206} \and
  S. Piacentini\thanksref{addr4,addr5}\orcidlink{0000-0002-1256-7149} \and
  D. Pierluigi\thanksref{addr6} \and
  D. Pinci\thanksref{addr10}\orcidlink{0000-0002-7224-9708} \and
  F. Renga\thanksref{addr10}\orcidlink{0000-0001-8129-8504} \and
  A. Russo\thanksref{addr6} \and
  G. Saviano\thanksref{addr6,addr12} \and
  P.A.O.C. Silva\thanksref{addr1}\orcidlink{0000-0002-1957-2274} \and
  N.J. Spooner\thanksref{addr13} \and
  R. Tesauro\thanksref{addr6}\orcidlink{0009-0006-0722-5896} \and
  S. Tomassini\thanksref{addr6}\orcidlink{0000-0001-7290-2028} \and
  D. Tozzi\thanksref{addr9,addr10}\orcidlink{0009-0001-9206-7354}
}

\thankstext{e1}{e-mail: melba.dastolfo@gssi.it}
\thankstext{e2}{e-mail: dho@lnf.infn.it}
\thankstext{e3}{e-mail: davide.fiorina@gssi.it}

\institute{
  LIBPhys; Department of Physics; University of Coimbra; 3004-516 Coimbra; Portugal \label{addr1}
  \and
  Dipartimento di Matematica e Fisica; Università Roma TRE; 00146; Roma; Italy \label{addr2}
  \and
  Istituto Nazionale di Fisica Nucleare; Sezione di Roma Tre; 00146; Rome; Italy \label{addr3}
  \and
  Gran Sasso Science Institute; 67100; L'Aquila; Italy \label{addr4}
  \and
  Istituto Nazionale di Fisica Nucleare; Laboratori Nazionali del Gran Sasso; 67100; Assergi; Italy \label{addr5}
  \and
  Istituto Nazionale di Fisica Nucleare; Laboratori Nazionali di Frascati; 00044; Frascati; Italy \label{addr6}
  \and
  ENEA Centro Ricerche Frascati; 00044; Frascati; Italy \label{addr7}
  \and
  Universidade Federal de Juiz de Fora; Faculdade de Engenharia; 36036-900; Juiz de Fora; MG; Brasil \label{addr8}
  \and
  Dipartimento di Fisica; Universit\`a di Roma Sapienza; 00185; Roma; Italy \label{addr9}
  \and
  Istituto Nazionale di Fisica Nucleare; Sezione di Roma; 00185; Roma; Italy \label{addr10}
  \and
  Universidade Estadual de Campinas -- UNICAMP; Campinas 13083-859; SP; Brazil \label{addr11}
  \and
  Dipartimento di Ingegneria Chimica; Materiali e Ambiente; Sapienza Universit\`a di Roma; 00185; Roma; Italy \label{addr12}
  \and
  Department of Physics and Astronomy; University of Sheffield; Sheffield; S3 7RH; UK \label{addr13}
  \and
  CERN, 1211 Geneva 23, Switzerland \label{addr14}
}

\date{Received: \today}

\maketitle

\begin{abstract}
The CYGNO project develops optical time projection chambers for
directional dark matter searches, using a He:CF$_4$ (60:40) gas
mixture at atmospheric pressure and a triple-GEM amplification
stage. We investigate the addition of 1--10\% HFO-1234ze as a
hydrogen-containing additive with low global warming potential.
The charge response and scintillation light yield are studied
with the MANGO prototype, complemented by secondary-scintillation
spectroscopy of the baseline mixture and mixtures containing
1\% and 2.5\% HFO. Increasing the HFO concentration shifts the
charge-response curves towards higher GEM voltages and strongly
suppresses the scintillation emission in both the ultraviolet
and visible bands. We use the measured optical response to
estimate the impact on projected spin-independent WIMP--nucleon
and spin-dependent WIMP--proton exclusion limits for a CYGNO-like
directional detector. Under the assumed detector-response model,
the higher optical energy threshold largely outweighs the
benefit of the additional light target, although a 1\% HFO
admixture retains a limited advantage for low-mass
spin-dependent interactions. An idealized fixed-threshold
benchmark is also considered to illustrate the potential of
the modified target composition without the optical threshold
penalty.
\end{abstract}

\section{Introduction}
\label{introduction}
The concept of dark matter (DM) was introduced almost one century ago to justify astrophysical observations indicating the presence of a diﬀerent kind of matter that does not interact with the electromagnetic force. Despite a growing body of evidence over the years, the true nature of dark matter remains undetermined, making its existence one of the most important open issues in fundamental physics, as described in \cite{bertone2005particle}.

Among the possible DM candidates, Weakly Interacting Massive Particles (WIMPs) have long been well motivated and extensively studied (\cite{mayet2016review}). However, current experimental and theoretical efforts increasingly focus on WIMP-like particles, which retain similar interaction phenomenology while extending beyond the standard WIMP paradigm. A well-established technique for investigating the DM problem is based on detecting these particles from our Galactic halo as they pass through the Earth. The basic idea is to measure the small energy deposited when such a particle scatters off a nucleus in a well-instrumented target volume. This technique, known as direct detection, forms the basis of
many experimental searches; a comprehensive review is provided
in \cite{Billard2022APPEC}.

The CYGNO experiment~\cite{amaro2022cygno} develops gaseous time
projection chambers (TPCs) with optical readout for directional
dark matter searches. The baseline detector operates at
atmospheric pressure with a He:CF$_4$ (60:40) gas mixture and a
triple-GEM amplification stage~\cite{SAULI1997531}. Ionisation
electrons drift towards the GEM stack, where avalanche
multiplication produces secondary scintillation
light~\cite{Morozov_2012}. Scientific cameras record the track
projection on the $X$--$Y$ plane, while photomultiplier tubes
(PMTs) measure the light-emission time profile, which provides
information on the relative extent of the track along the
drift direction, $\Delta z = v_d\Delta t$, where $v_d$ is the
electron drift velocity. This gives the relative longitudinal extent, not the absolute interaction depth. Reconstructing recoil directions
can help distinguish the anisotropic Galactic dark matter
signal from backgrounds~\cite{mayet2016review}. The sensitivity
study presented in Sect.~\ref{sec:limits} adopts a baseline
electron-equivalent energy threshold of 0.5~keV$_{ee}$,
following the projected scenario of
Ref.~\cite{amaro2022cygno}; this value is a sensitivity-model
input rather than a threshold measured in the present campaign.

Despite extensive efforts in the direct detection of WIMPs, no signals have been observed so far, resulting in the exclusion of a large portion of the possible WIMP cross section--mass parameter space. That has renewed the attention to WIMP masses below \SI{1}{\giga\electronvolt/c^2}, a region still theoretically well-founded and largely unexplored to these days (\cite{zurek2024}). To extend CYGNO's sensitivity down to this lower-mass regime, the addition of a lighter target gas becomes essential. The most suitable option is a hydrogen-based gas, which offers favorable kinematic matching to WIMP-like DM candidates with $\mathcal{O}$(GeV/$c^2$) masses and provides sensitivity to spin-dependent (SD) DM-proton couplings. Previous studies conducted by our collaboration focused on hydrocarbons, in particular isobutane (\cite{AMARO2024138759}) and methane (\cite{methane}), as they have proven to be an excellent choice as additives to the base He:CF$_{4}$ mixture. Unfortunately, the gas system planned for the CYGNO demonstrator is not currently suitable for flammable gases; therefore, an alternative has to be pursued. The choice fell upon 1,3,3,3-Tetrafluoropropene, commercially known as HFO-1234ze (hereafter HFO), a hydrofluoroolefin refrigerant with very low Global Warming Potential (GWP$_{100}$ $\approx$ 7). HFO has been extensively studied as part of the transition process to eco-friendly gas mixtures for several gaseous detector technologies (\cite{Abbrescia2024_ECOGasGIFPP}, \cite{Benussi:2015qqa}, \cite{ECOgasGIFPP2025_EPJPlus}).
The reduction of the mass-weighted GWP relative to the baseline
mixture is reported in Table~\ref{tab:gwp_mixtures}.

\begin{table}[t]
  \centering
\caption{Calculated mass-weighted 100-year GWP of
He:CF$_4$ (60:40) with HFO-1234ze admixtures.
HFO concentrations are volume fractions in the final
mixture. Reductions are relative to the baseline mixture
on an equal-mass basis.
Constituent GWP values are referenced to
\cite{IPCC_AR6_WGI_Chapter7_2021}.}
  \label{tab:gwp_mixtures}
  \begin{tabular}{ccc}
    \hline\noalign{\smallskip}
    HFO-1234ze(E) [\%] & GWP$_{100}$ & Reduction [\%] \\
    \noalign{\smallskip}\hline\noalign{\smallskip}
    0.0  & 6909 & 0.0  \\
    1.0  & 6703 & 3.0  \\
    2.5  & 6410 & 7.2  \\
    5.0  & 5958 & 13.8 \\
    7.5  & 5545 & 19.7 \\
    10.0 & 5168 & 25.2 \\
    \noalign{\smallskip}\hline
  \end{tabular}
\end{table}

HFO was also selected under the working hypothesis that electron-impact fragmentation (dissociative ionization/attachment) can produce abundant \(\mathrm{CF_3^+}\) and neutral \(\mathrm{CF_3}\) (possibly in excited states, \(\mathrm{CF_3}^*\)), whose radiative de-excitation could increase the light yield in our detector, thereby motivating the tests (\cite{Lyshchuk2025_PhysScr}).
In this work, we investigate the addition of 1.0--10.0\% HFO
to the baseline He:CF$_4$ mixture. We characterise the charge
response and optical light yield with MANGO, measure
secondary-scintillation spectra for selected mixtures, and
use these results to estimate the impact on projected
spin-independent and spin-dependent dark matter exclusion
limits.
\section{Experimental setup}
\label{exp_setup}
The detector selected for this measurement campaign was MANGO, one of the smallest prototypes developed by the CYGNO Collaboration. Given its compact size, it was the ideal option for characterizing the gain response to the HFO admixture. These
measurements were performed in the Gas Detector Development (GDD) laboratory at CERN.

For the spectroscopic measurement of the secondary scintillation, we relied on an apparatus already available in the GDD laboratory. This setup is based on the same concept as MANGO, a GEM-based TPC, and is integrated with an X-ray generator and a fiber-coupled CCD spectrometer.

In both setups, the gas mixtures were prepared and operated at atmospheric pressure and room temperature, following the subsequent method: the detectors were first filled with HFO up to the desired fraction of the total pressure, and then completed with He:CF$_4$ in a fixed 60:40 volume ratio.

The setups are described in detail in the following subsections.

\subsection{MANGO}
\begin{figure}[h]
  \centering
  \includegraphics[width=\linewidth]{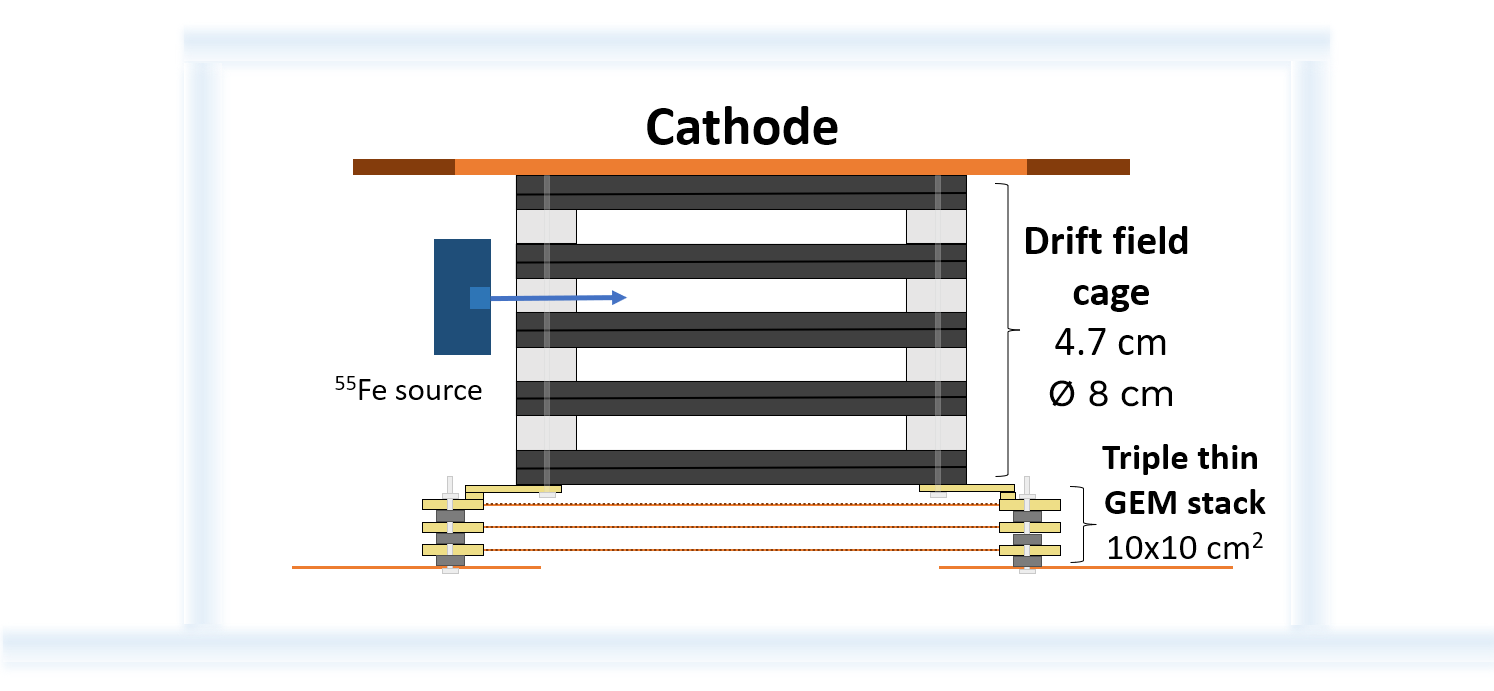}
  \caption{Schematic view of the MANGO detector (not to scale).}
  \label{MANGO}
\end{figure}
The schematic of MANGO's experimental setup is shown in Fig.~\ref{MANGO}. It reproduces the CYGNO baseline concept on a small scale: a TPC with 4.7~cm drift gap and a cylindrical active volume of 3.7~cm radius, equipped with a standard triple 10$\times$10~cm$^2$ GEM stack. Each GEM consists of a 50~$\mu$m thick Kapton foil coated on both sides with 5~$\mu$m thick Cu and perforated with biconical holes, 70~$\mu$m in diameter in the copper layers and 50~$\mu$m diameter in Kapton, in a hexagonal pattern with \SI{140}{\micro\meter} pitch. 
The detector was irradiated with a $^{55}$Fe X-ray source,
using a nominal photon energy of 5.9~keV for the
charge-response and optical light-yield measurements
reported below.
For the measurement presented in this study, the three GEMs were powered through a CAEN A1515TG board, which provides independent floating negative voltages to the GEM electrodes, while the cathode was supplied with a negative high voltage by one channel of a CAEN N1570 module. The second channel of the same module fed the last ring of the field cage, whose intermediate rings were biased through a resistive divider chain.
The voltage drop across GEM $i$ is denoted by $\Delta V_i$.
The GEM-voltage variable used in
Figs.~\ref{electron_gain} and~\ref{light_yield} denotes
the voltage drop across each GEM foil.

The drift field was 1~kV/cm.
The transfer fields between GEM1--GEM2 and GEM2--GEM3
were 2.5~kV/cm  across gaps of
2~mm.

The TPC is separated from the external environment by two distinct layers: an inner airtight acrylic vessel that contains the gas and houses the detector elements, and an outer light-tight enclosure made of 3D-printed black plastic. A thin Mylar window with optical transmission greater than 0.9 separates the gas volume from the external optical readout. The window is located \SI{5}{\milli\meter} from the last GEM. The optical readout comprises a Hamamatsu Orca Quest qCMOS camera \cite{hamamatsu_orca_quest2} and one PMT (Hamamatsu R1635 \cite{hamamatsu_r1635}), placed next to the camera and used for triggering the acquisition. The camera is characterized by 4096$\times$2304 pixels, each with a read noise of 0.27 electron root mean square (RMS), and a dimension of 4.6$\times$4.6~$\mu$m$^2$. It is equipped with an EHD lens \cite{ehd_f085_lens} with focal length $f=25$~mm and f-number $N=f/D=0.85$,
where $D$ is the entrance-pupil diameter. In this configuration, the optical system images an area of 14.2$\times$8.0~cm$^2$ focused on the GEM3 plane with an effective pixel size of 35$\times$35~$\mu$m$^2$ and a geometrical light-collection fraction
$A_{\mathrm{geom}}=\Omega_{\mathrm{lens}}/(4\pi)
=1.18\times10^{-3}$, where $\Omega_{\mathrm{lens}}$ is
the solid angle subtended by the lens at the emitting region.
As the longitudinal resolution of the track position is not necessary for the characterization of the light yield of the mixture, this information will not be explored in this paper.

\subsection{Spectroscopic measurement setup}
\label{spectroscopy_setup}
\begin{figure*}[t]
  \centering
  \includegraphics[width=0.95\textwidth]{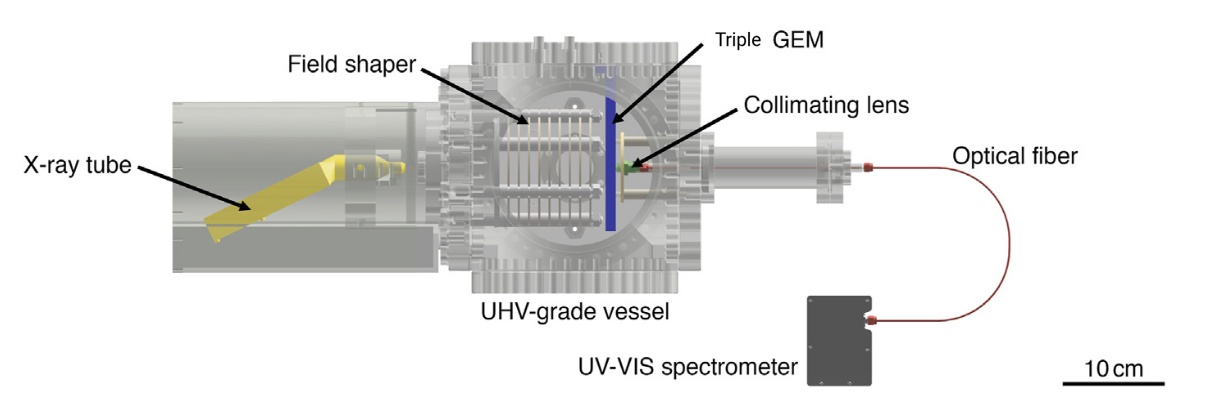}
  \caption{Spectroscopy setup for secondary
  scintillation measurements.}
  \label{tube}
\end{figure*}

Figure~\ref{tube} shows the secondary-scintillation spectroscopy setup employed in this study.

The concept behind this setup is as follows: an X-ray tube irradiates a TPC-like detector from the left, passing through a thin Al foil window and a mesh cathode. A portion of the X-rays emitted is absorbed by the gas in the drift region, generating primary electrons. These electrons are then guided towards the amplification stage by a weak electric field, where they undergo charge avalanche multiplication in the strong electric field within the GEM holes. During this process, scintillation light is produced and collected by a collimating lens positioned beneath the GEM stack. The light is then transmitted through a feedthrough to a UV--VIS spectrometer\footnote{Spectrometer: OceanOptics Flame CCD spectrometer. Fiber: 600~$\mu$m core MM optical fiber with UV-VIS-NIR transparency. Collimating lens: 74-UV lens with 5~mm diameter active area} located outside the detector volume. The spectrometer and optical system, consisting of the collimating lens, optical fibers, and feedthrough, as well as the CCD sensitivity, were calibrated using a deuterium-halogen calibration lamp to ensure accurate spectral measurements.

The TPC configuration used in this study featured a 10~cm diameter and 10~cm-long field shaper. The GEM stack was arranged such that the `top' electrode of the first GEM (facing the cathode) was grounded, while the remaining electrodes were individually biased using a high-voltage power supply. The current collected at the bottom electrode of the third GEM (facing the anode) was continuously monitored. This measurement is essential because it provides the basis for normalizing the measured spectra, allowing direct comparison across different operating conditions.

For the measurements presented here, the X-ray tube was
operated at 15~kV. The expected source spectrum contains
a bremsstrahlung continuum with an endpoint energy of
15~keV, rather than monochromatic 15~keV irradiation.
The tube used an Ag anode.
The exposure time for each acquisition was 30~s. Multiple spectra were recorded and averaged for each gas mixture to reduce statistical fluctuations. Background spectra were also acquired with the X-ray tube turned off and the detector powered down and subsequently subtracted from the signal to eliminate contributions unrelated to secondary scintillation, including the offset value and the dark current of the CCD spectrometer.

Further information on the experimental setup can be found in \cite{Brunbauer:2933707}.

\section{Effective charge yield}
\label{sec:charge}

\begin{figure}[h]
  \centering
  \includegraphics[width=\linewidth]{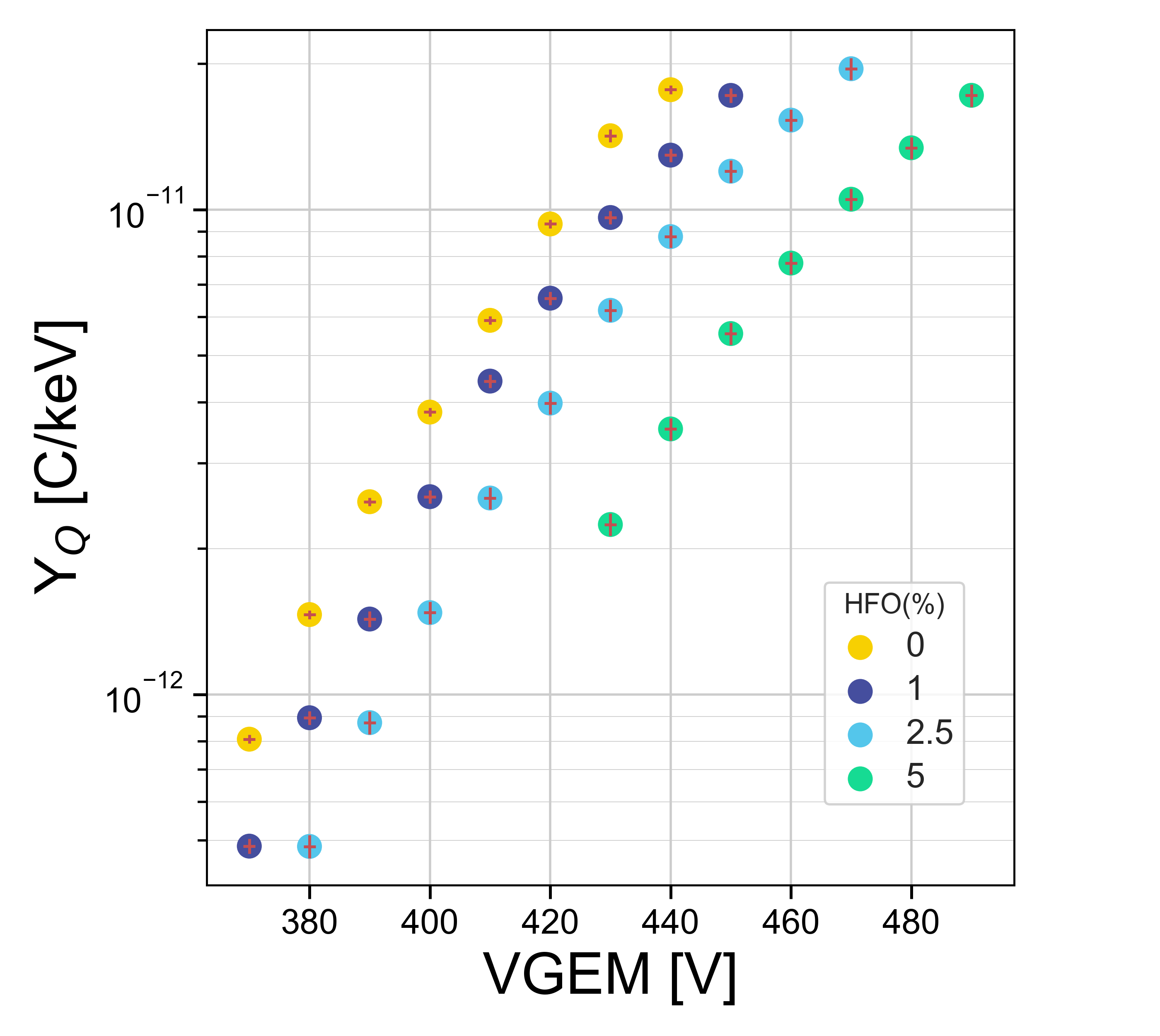}
  \caption{Effective charge yield $Y_Q$, in C/keV,
  as a function of GEM voltage for HFO admixtures
  of 0, 1, 2.5, and 5\%.
  For the 5\% HFO mixture, no points are shown at
  low GEM voltages because the signal current was
  below the measurement sensitivity.}
  \label{electron_gain}
\end{figure}

To assess the charge response to HFO admixtures,
we measured the effective charge yield of the MANGO TPC.
This quantity expresses the collected avalanche charge
per unit nominal deposited energy without assuming
a $W$-value for each mixture.

The charge signal was collected at the bottom electrode
of the third GEM. The source-induced current magnitude
is defined as
$I_{\mathrm{sig}}=|I_{\mathrm{on}}-I_{\mathrm{off}}|$,
where $I_{\mathrm{on}}$ and $I_{\mathrm{off}}$ are
measured with the source exposed and blocked,
respectively. For each mixture, the event rate was
measured as a function of the GEM voltage. The rate
increases and then reaches a plateau, whose value,
$R_{\max}$, is of order 10~kHz.

Using the nominal $^{55}$Fe photon energy
$E_{\mathrm{Fe}}=5.9$~keV, the effective charge yield is
\begin{equation}
 Y_Q=\frac{I_{\mathrm{sig}}}
 {R_{\max}E_{\mathrm{Fe}}},
 \qquad [Y_Q]=\mathrm{C/keV}.
 \label{eq:charge_yield}
\end{equation}
This differs from the dimensionless effective electron
gain, which would be $G_{\mathrm{eff}}=Y_QW/e$ when
$W$ is expressed in keV per electron--ion pair and
$e$ is the elementary charge. Since $W$ is not known
for all the He:CF$_4$:HFO mixtures, the comparison
below is expressed in terms of $Y_Q$.

The current was measured with an ammeter%
\footnote{Keithley Model 6487 Picoammeter/Voltage Source}.
At each voltage setting, measurements with the source
blocked and exposed were used to determine the
source-induced current. The event rate was measured
through a standard NIM chain composed of a preamplifier,
shaper, discriminator, and scaler.

Figure~\ref{electron_gain} shows that increasing the
HFO concentration reduces $Y_Q$ at a fixed GEM voltage,
requiring higher voltages to recover a comparable
charge yield. The highest voltage point for each
mixture was set 10~V below the observed discharge
onset on each GEM. This onset was determined
independently for each mixture by gradually
increasing the GEM voltages.
The maximum effective charge yields reached in the
reported 0--5\% HFO scans are comparable.

To quantify the relative loss at a fixed GEM voltage $V$,
we define $\ell_Q(x,V)=1-Y_Q(x,V)/Y_Q(0,V)$, where $x$ is
the HFO fraction and $x=0$ denotes the baseline mixture.
Reading the plotted central values in Fig.~\ref{electron_gain},
the losses are approximately 32\%, 56\%, and 84\% at 430~V,
and 27\%, 50\%, and 80\% at 440~V, for 1\%, 2.5\%, and
5\% HFO, respectively. These are the two GEM-voltage
settings for which measurements of all four mixtures are
available.

\section{Effective light yield}
\label{sec:lightY}
Because the CYGNO detection technique relies on an optical readout, an accurate estimation of the effective light yield is particularly important to determine the potential of new gas mixtures.

All measurements presented here were performed with the MANGO prototype irradiated with the ${}^{55}$Fe source, using the qCMOS camera. The images are processed with the standard reconstruction code developed by the CYGNO collaboration, which selects the pixels corresponding to each signal (soft electrons, muons, alpha tracks) and estimates features such as energy deposited, size of the cluster, and more. A detailed description of the operation of the algorithm can be found in \cite{dho2024ely,pains2023idbscan,dimarco2020identification}. The acquired data are analyzed by the reconstruction algorithm, the signals are identified, and a spectrum of the \textit{Integral} of the clusters is obtained. The \textit{Integral} is defined as the sum of the intensities of all pixels belonging to a cluster, representing a quantity proportional to the energy deposited in the gas. Therefore, the effective light yield (LY) is taken as
the mean \textit{Integral} of the measured clusters
when the detector is irradiated with the
${}^{55}$Fe source. It is expressed in camera analog-to-digital converter (ADC) counts.

\begin{figure}[h]
  \centering
  \includegraphics[width=\linewidth]{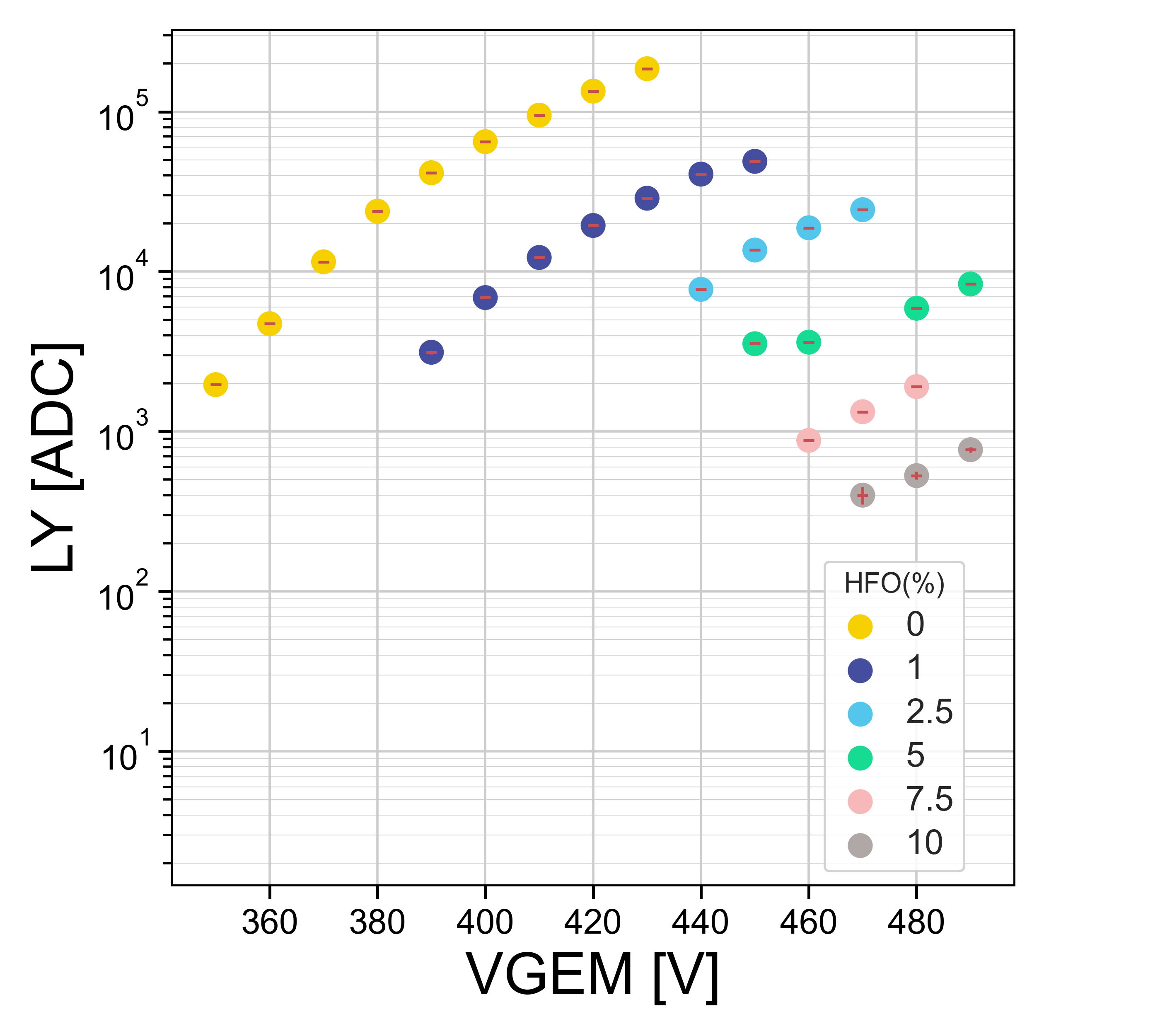}
\caption{Effective light yield (LY) as a function of GEM voltage
for the different HFO percentage additions to the baseline
He:CF$_4$ mixture.}
  \label{light_yield}
\end{figure}

\begin{figure}[h]
  \centering
  \includegraphics[width=\linewidth]{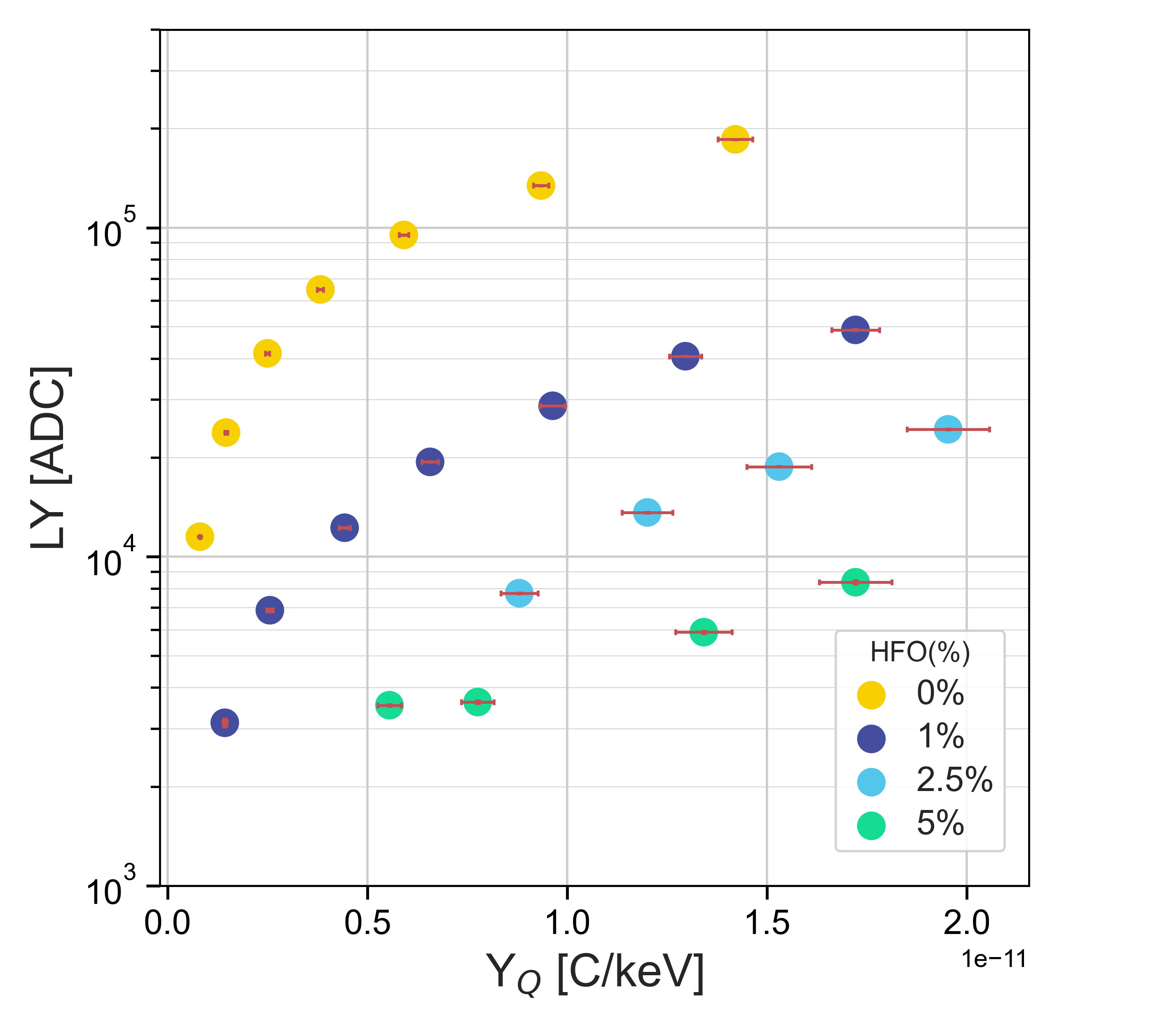}
\caption{Effective light yield (LY) as a function of effective charge yield $Y_Q$ for the different HFO percentage additions to the baseline He:CF$_4$ mixture.}
  \label{light_yield_vs_gain}
\end{figure}

The results are presented in Fig.~\ref{light_yield} and Fig.~\ref{light_yield_vs_gain}, showing the light yield as a function of the GEM
biasing voltage and of the effective charge yield
$Y_Q$, respectively. As is clearly visible, the increase in HFO concentration prevents achieving the same LY as the standard baseline, even when the GEM voltage is raised. This is an evident indication of the light quenching effect introduced by HFO. Contrary to the initial working hypothesis, no light-yield enhancement from a possible radiative de-excitation of CF$_3^*$ fragments was observed.


\section{Secondary scintillation spectra}
To gain a more complete picture of the light emission produced by different HFO admixtures with the baseline gas of the CYGNO experiment, it is also necessary to measure their secondary scintillation spectra.

The detector, described in Sect.~\ref{spectroscopy_setup}, was operated at room temperature and pressure.

\begin{figure}[h]
  \centering
  \includegraphics[width=\linewidth]{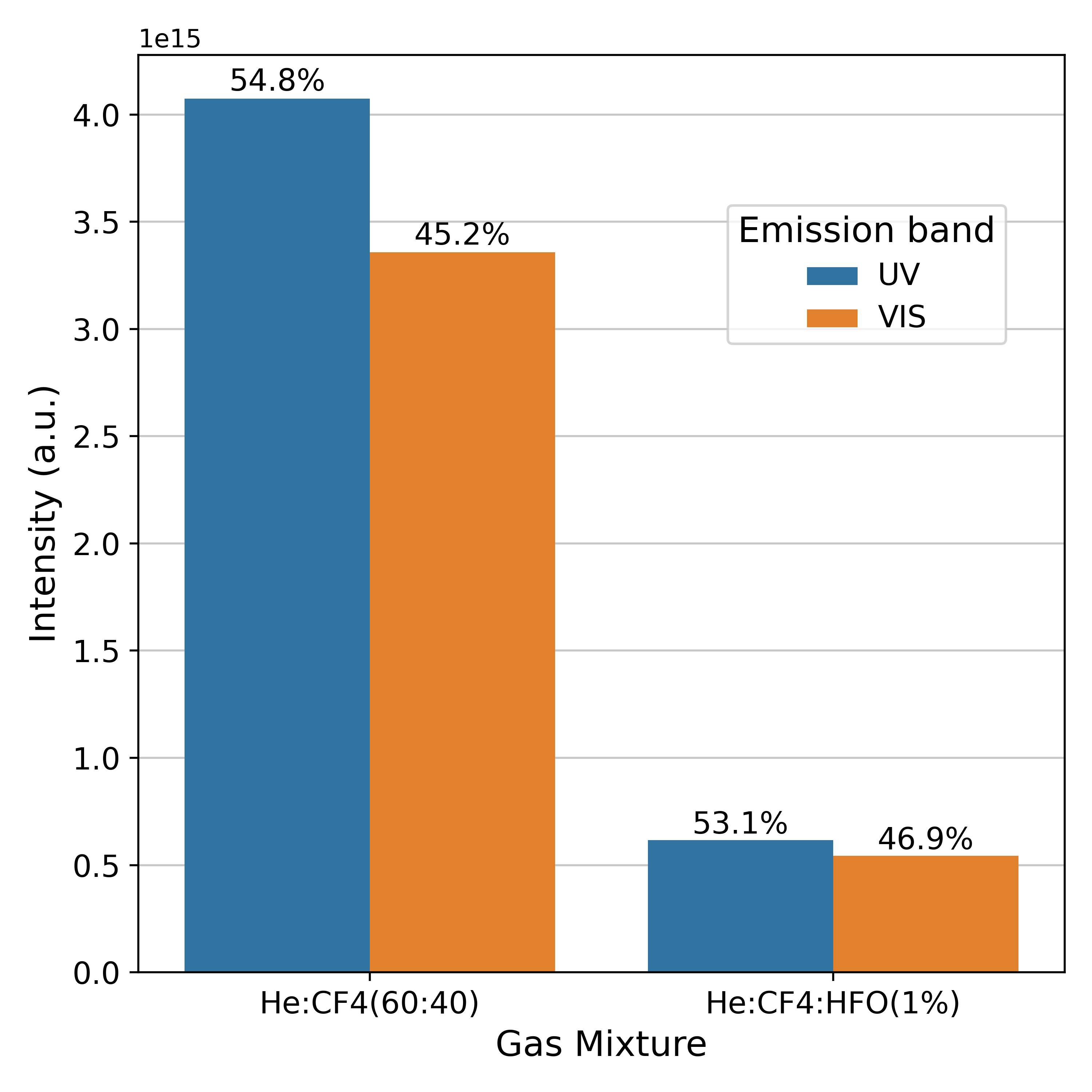}
  \caption{Integrated UV and VIS secondary-scintillation
intensities for the baseline He:CF$_4$ mixture and
the mixture containing 1\% HFO.
The spectra are normalized to the collected current
and integrated over 200--400~nm (UV) and
400--750~nm (VIS).
The percentages above the bars give each band's
contribution to the sum of the two integrals
within the same mixture.}
  \label{uv_vis}
\end{figure}

Acquiring data at the maximum stable gain with this setup required careful selection of both the GEM voltages and the X-ray tube current. Therefore, each spectrum was recorded only after it could be consistently reproduced multiple times, even under slight variations in the operating conditions.
\begin{figure}[h]
  \centering
  \includegraphics[width=\linewidth]{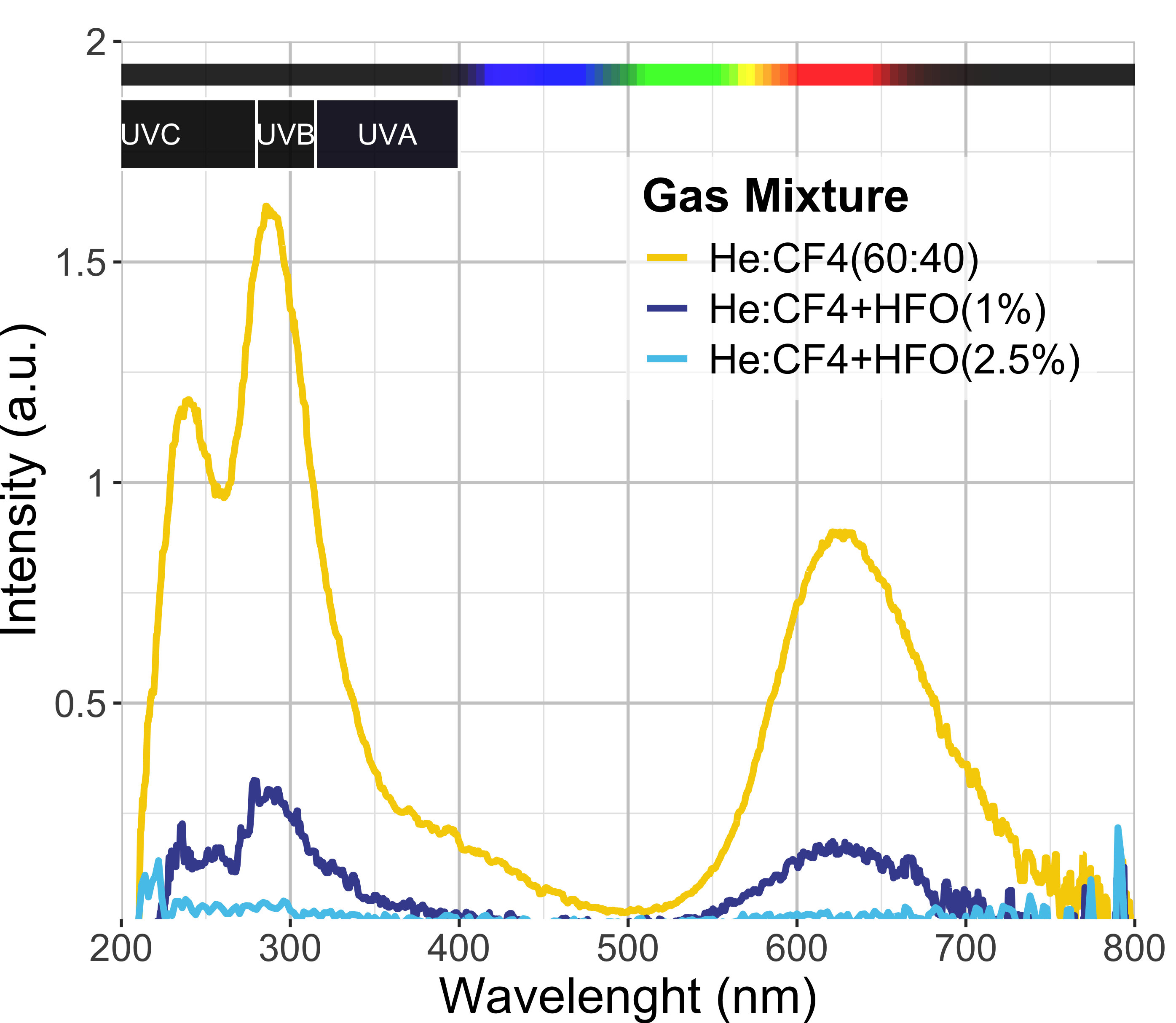}
  \caption{Secondary scintillation spectra of He:CF$_4$, He:CF$_4$ + HFO(1\%) and He:CF$_4$ + HFO(2.5\%) gas mixtures at 1~bar: scintillation intensity normalised to collected current.}
  \label{spectra}
\end{figure}
In Fig.~\ref{spectra} three different spectra, normalized to the collected electron current, are presented: the baseline gas mixture (yellow line), and two other mixtures with HFO additives, the former with 1\% HFO (dark blue) and the latter with 2.5\% HFO (light blue). It is important to note that the signal-to-noise ratio of the spectrometer decreases near the edges of its sensitivity range (200--800~nm), leading the acquisition software to apply corrections that may amplify the statistical fluctuations in that range. The plot clearly shows that the HFO leads to a significant light yield quenching, both in the visible and UV bands. This strong suppression is the reason why other HFO-richer gas mixtures were not investigated any further, as they were unable to provide sufficient light emission. For each band $b$, we define the relative light-yield loss as
$\ell_b(x)=1-J_b(x)/J_b(0)$, where $J_b$ is the integral of
the background-subtracted, response-corrected and
current-normalized spectrum. The integration ranges are
200--400~nm for UV and 400--750~nm for VIS.
Reading the band integrals plotted in Fig.~\ref{uv_vis},
the relative losses for 1\% HFO are approximately
85\% in the UV band and 84\% in the VIS band.
Figure~\ref{uv_vis} shows the band integrals and the UV/VIS
fractions within each mixture. The 2.5\% mixture is not
included in this comparison because its low light yield
produces large uncertainties.

\section{Implications for dark matter sensitivity}
\label{sec:limits}
To estimate the effect of introducing an HFO gas in the CYGNO mixture, the exclusion limits for WIMP-nucleon spin-independent (SI) and WIMP-proton spin-dependent (SD) interaction are studied. A Bayesian simulation technique is employed to assess the expected limits for a CYGNO-like directional detector. This method is based on simulating datasets of pure background samples on the 2D angular Galactic coordinate plane, which are fitted with a likelihood function that includes both background and standard WIMP signal components. Afterwards, exploiting the Bayes theorem, the posterior probability on the number of signal events can be calculated and the limit for each DM mass is taken as the 90\% credible interval (C.I.) of the resulting curve. This method is described in detail in \cite{amaro2022cygno}. The experimental elements included in this study are the multielement composition of the gas mixture, the energy threshold, the effect of the angular resolution (kept constant for each energy to 30$\times$30~deg$^2$), and the quenching factor of the recoiling nuclei. The detector concept considered for the analysis is a generic
CYGNO-like directional detector of 30~m$^3$ operated for
3 years, corresponding to an exposure of 153~kg$\cdot$yr.
The reference energy range of interest is 0.5--300~keV$_{ee}$,
with the mixture-dependent lower thresholds specified in
Table~\ref{tab:thres}.
A residual background of $10^4$ events per year,
corresponding to $3\times10^4$ events over the full exposure,
is used as a common benchmark for all mixtures.
This assumption does not model the dependence of the residual
background on the threshold or HFO concentration. Following the studies of the TREX and Borexino experiments\cite{Trex,BENZIGER2007509}, employing clean materials for field cage, cathode, GEM and vessel can reasonably point at background levels of $10^4$--$10^5$ events per year prior any background discrimination which is expected to be extremely effective in imaging detectors\cite{Ghrear_2021}. Thus, while the present
measurements do not establish that this background level can be achieved, it makes sense to employ this value as a sensitivity-model input.

As shown in Sect.~\ref{sec:lightY}, the increase of the content of HFO gas in the mixture resulted in a reduction of the maximum effective light yield reached. As a result, the energy threshold of the detection is increasing with an inverse proportional relation. The angular response is held fixed in this comparison; a possible degradation of directional reconstruction with reduced light yield~\cite{Billard_2012} is not included. For the standard He:CF$_4$ mixture, the threshold is taken as 0.5~keV$_{ee}$ as described in \cite{amaro2022cygno}. To evaluate the energy threshold of each HFO-containing mixture, the maximum light yields reached in Sect.~\ref{sec:lightY} are compared with the one attained with no HFO. Table~\ref{tab:thres} shows the value of the computed energy threshold for all the HFO data sets.
\begin{table}[!t]
  \centering
  \caption{Energy threshold in keV$_{ee}$ for all the HFO mixture data sets analysed in this work. Uncertainties on the energy threshold are of 2\% coming from the light yield analysis.}
  \label{tab:thres}
  \begin{tabular}{cc}
    \hline\noalign{\smallskip}
    HFO (\%) & $E_\mathrm{thr}$ (keV$_{ee}$) \\
    \noalign{\smallskip}\hline\noalign{\smallskip}
    0   & 0.5 \\
    1   & 1.1 \\
    2.5 & 4.5 \\
    5   & 13  \\
    7.5 & 51  \\
    10  & 127 \\
    \noalign{\smallskip}\hline
  \end{tabular}
\end{table}
The optical-readout limits shown below are evaluated for
0, 1, and 2.5\% HFO. Mixtures with 5\% HFO or more,
which have higher scaled thresholds, are not included
in this comparison.

\begin{figure}[!t]
  \centering
  \includegraphics[width=\linewidth]{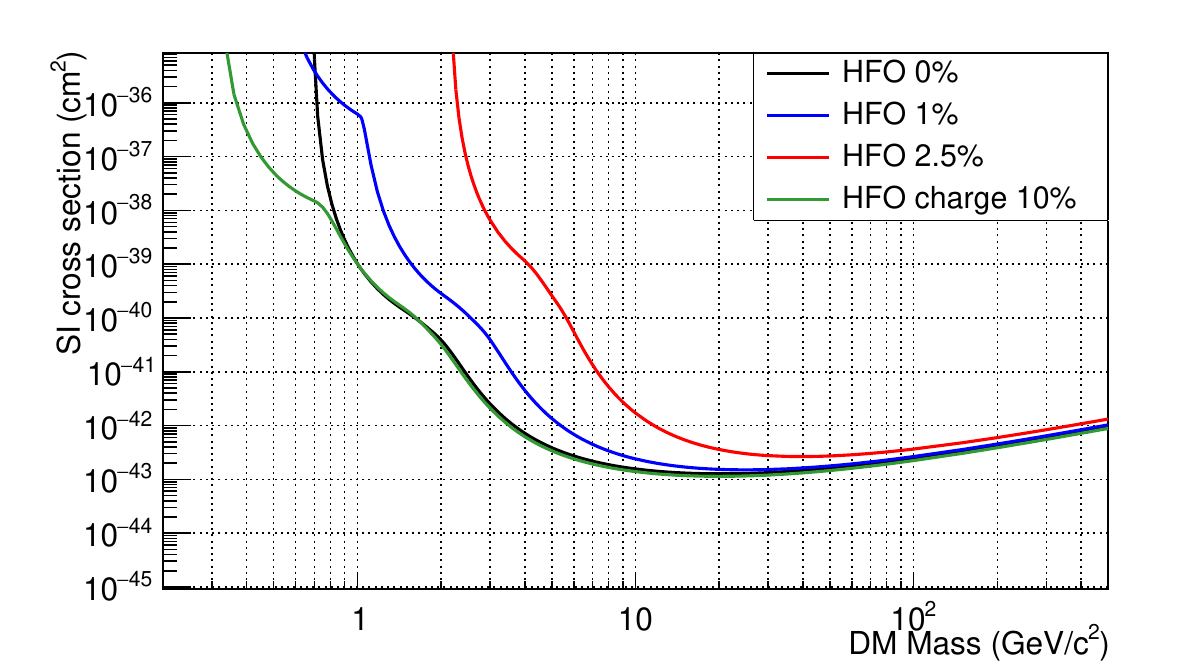}
  \includegraphics[width=\linewidth]{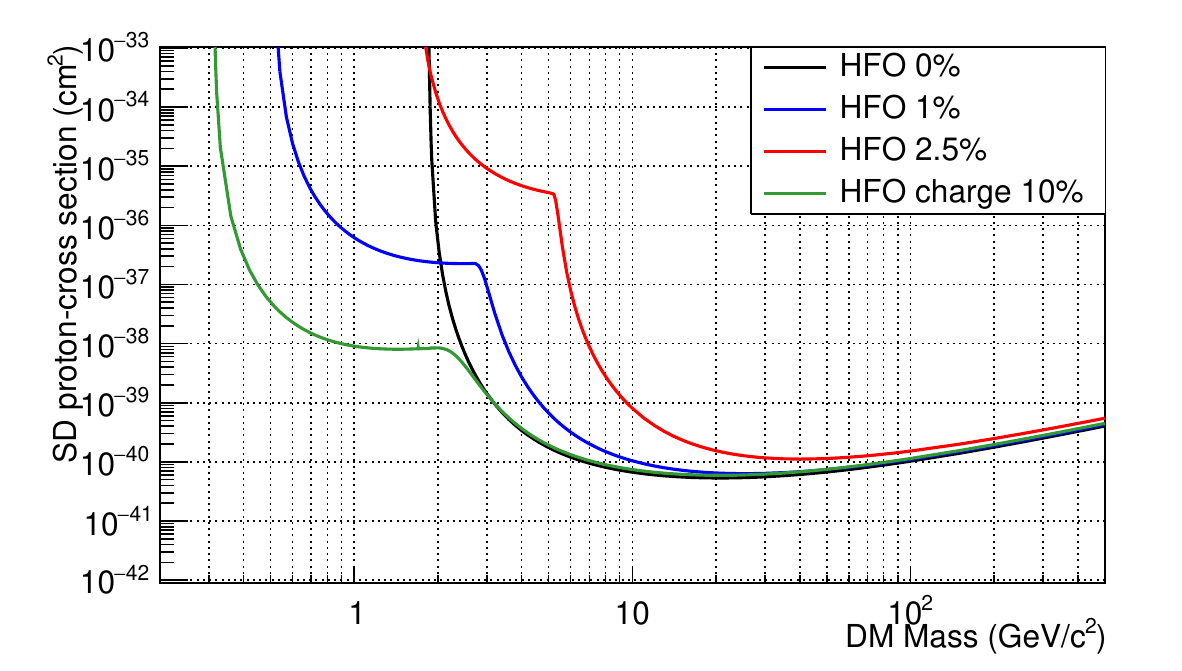}
  \caption{90\% C.I. for the SI WIMP to nucleon (top) and SD WIMP to proton (bottom) cross section as a function of the DM mass for a CYGNO-like detector with 153~kg$\cdot$yr exposure and $3\cdot10^4$ total background events. Gas mixtures with different HFO contents are displayed. The green line represents a hypothetical 10\% HFO charge-readout detector with a 0.5~keV$_{ee}$ threshold and the same assumed angular response and residual background.}
  \label{SI}
\end{figure}

The 90\% C.I. limits for the SI and SD interactions are displayed in Fig.~\ref{SI}, top and bottom panels respectively. The mixture with 2.5\% of HFO has much poorer limits in both SI and SD cases, with no advantage gained with the addition of hydrogen in the gas. The mixture with 1\% offers improvements in the SD case, allowing to achieve sensitivity down to about 0.5~GeV/$c^2$ DM mass. However, this is obtained at the cost of sensitivity at intermediate masses and especially loss of sensitivity in the whole mass range in the SI analysis.

We also calculate the SI and SD limits for a hypothetical 10\% HFO detector with charge readout, keeping the threshold at 0.5~keV$_{ee}$ and the angular response and residual background at their reference values. These are benchmark assumptions, not consequences of the measured charge yield. A possible implementation would extract avalanche electrons below GEM3 onto a segmented anode: channel positions would provide transverse information and time-resolved signals would provide relative longitudinal information. Background rejection would depend on the readout granularity and noise and would have to be established for that implementation. Examples of charge readout can be found in other directional detector studies\cite{Ghrear_2021,NAMAI2025170145,Beaufort_2024}. Under the stated assumptions, the green curves in Fig.~\ref{SI} illustrate sensitivity down to about 0.3~GeV/$c^2$.

\section{Summary and conclusions}
In this work, we investigated the performance of He:CF$_4$ (60:40) mixtures enriched with controlled fractions of HFO-1234ze, used here as a practical route to introduce hydrogen targets in a CYGNO-like optical TPC while keeping a low GWP additive. The charge yield and optical response were measured with the MANGO prototype and complemented by dedicated secondary scintillation spectroscopy with a GEM based setup, enabling a quantitative assessment of the performance trade-offs alongside the expected sensitivity benefits from the added hydrogen content.

The effective charge-yield measurements show comparable maximum yields for the reported HFO admixtures up to 5\%, with higher GEM voltages needed as the HFO concentration increases. This behaviour is consistent with increased avalanche quenching. In contrast, the optical response degrades rapidly with increasing HFO content: the effective light yield measured with the qCMOS readout decreases significantly and cannot be recovered by raising the GEM voltage within safe operating conditions. The spectroscopic measurements confirm a strong suppression of secondary scintillation over both the UV and visible ranges already at the percent level in HFO concentration.

We evaluated the impact of these effects on projected dark matter exclusion limits for a CYGNO-like directional detector using a Bayesian approach. While introducing hydrogen through HFO can, in principle, enhance sensitivity to spin-dependent DM-proton interactions at low masses, the increased energy threshold driven by light quenching dominates the performance for optical-readout operation. As a result, only percent-level HFO admixtures may retain a limited advantage in the low-mass spin-dependent regime, at the price of reduced sensitivity at intermediate and higher masses. The hypothetical charge-readout comparison illustrates the potential of these hydrogen-containing mixtures if a low threshold and adequate directional reconstruction and background rejection can be maintained.

Overall, these results provide quantitative guidance for the choice of sustainable gas mixtures in CYGNO-like detectors and motivate further R\&D on eco-friendly hydrogenated additives and alternative readout strategies that mitigate scintillation quenching.

\section*{Acknowledgements}
This project has received funding under the European Union's Horizon 2020 research and innovation programme from the European Research Council (ERC) grant agreement No 818744.

\bibliographystyle{spphys}
\bibliography{reference}

@article{Abbrescia2024_ECOGasGIFPP,
  author       = {Abbrescia, M. and Aielli, G. and Aly, R. and others},
  title        = {High-rate tests on resistive plate chambers operated with eco-friendly gas mixtures},
  journal      = {European Physical Journal C},
  year         = {2024},
  volume       = {84},
  pages        = {300},
  doi          = {10.1140/epjc/s10052-024-12545-8},
  url          = {https://doi.org/10.1140/epjc/s10052-024-12545-8},
  received     = {2023-10-25},
  accepted     = {2024-02-08},
  published    = {2024-03-22}
}

@article{ECOgasGIFPP2025_EPJPlus,
  author = {Quaglia, L. and Ramos, D. and others},
  title        = {Preliminary results on the long-term operation of RPCs with eco-friendly gas mixtures under irradiation at the CERN Gamma Irradiation Facility},
  journal      = {Eur. Phys. J. Plus},
  year         = {2025},
  volume       = {140},
  pages        = {40},
  doi          = {10.1140/epjp/s13360-024-05773-0},
  url          = {https://doi.org/10.1140/epjp/s13360-024-05773-0},
  date         = {2025-01-15}
}

@techreport{Benussi:2015qqa,
      author        = "Benussi, L. and Bianco, S. and Ferrini, M. and Passamonti,
                       L. and Pierluigi, D. and Piccolo, D. and Russo, A. and
                       Saviano, G.",
      title         = "{A study of HFO-1234ze (1,3,3,3-Tetrafluoropropene) as an
                       eco-friendly replacement in RPC detectors}",
      archivePrefix = "arXiv",
      eprint        = "1505.01648",
      reportNumber  = "INFN-REPORT-INFN-14-14-LNF",
      year          = "2015",
      url           = "https://cds.cern.ch/record/2014841",
      note          = "15 pages, 20 figures",
}

@article{Lyshchuk2025_PhysScr,
  author       = {Lyshchuk, Hlib and Metting van Rijn, Marnik and Paul, Anirban and Kočišek, Jaroslav and Ranković, Miloš and Nag, Pamir and Fedor, Juraj},
  title        = {Electron-induced dissociation of HFO1234ze: dissociative electron attachment and dissociative ionization},
  journal      = {Physica Scripta},
  year         = {2025},
  volume       = {100},
  number       = {5},
  pages        = {055409},
  doi          = {10.1088/1402-4896/adcb77},
  url          = {https://doi.org/10.1088/1402-4896/adcb77},
  publisher    = {IOP Publishing},
  note         = {Focus on the Fundamentals and Applications of Electron Attachment and Ion–Molecule Processes},
  date         = {2025-04-25}
}

@article{Brunbauer:2933707,
      author        = "Brunbauer, F M and Amedo, P and Flöthner, K J and
                       Gonzalez Diaz, D and Janssens, D and Leardini, S and
                       Lisowska, M and Müller, H and Oliveri, E and Orlandini, G
                       and Pfeiffer, D and Ropelewski, L and Sauli, F and
                       Samarati, J and Scharenberg, L and van Stenis, M and
                       Veenhof, R",
      title         = "Primary and secondary scintillation of CF4-based mixtures
                       in low-pressure gaseous detectors. Primary and secondary
                       scintillation of CF4-based mixtures in low-pressure gaseous
                       detectors",
      journal       = "Front. Detect. Sci. Tech.",
      volume        = "3",
      pages         = "1561739",
      year          = "2025",
      url           = "https://cds.cern.ch/record/2933707",
      doi           = "10.3389/fdest.2025.1561739"
}

@article{Billard_2012,
	doi = {10.1103/physrevd.85.035006},
	
	url = {https://doi.org/10.1103/PhysRevD.85.035006},
	
	year = 2012,
	month = {feb},
	
	publisher = {American Physical Society ({APS})},
	
	volume = {85},
	
	number = {3},
	
	author = {J. Billard and F. Mayet and D. Santos},
	
	title = {Assessing the discovery potential of directional detection of dark matter},
	
	journal = {Physical Review D}
}

@article{AMARO2024138759,
title = {Secondary scintillation yield from GEM electron avalanches in Image 1-Image 2 and Image 1-Image 2-isobutane for CYGNO — Directional Dark Matter search with an optical TPC},
journal = {Physics Letters B},
volume = {855},
pages = {138759},
year = {2024},
issn = {0370-2693},
doi = {https://doi.org/10.1016/j.physletb.2024.138759},
url = {https://www.sciencedirect.com/science/article/pii/S0370269324003174},
AUTHOR = {Amaro, F. D. and Baracchini, E. and Benussi, L. and Bianco, S. and Capoccia, C. and Caponero, M. and Cardoso, D. S. and Cavoto, G. and Cortez, A. and Costa, I. A. and {D'Imperio}, G. and Dané, E. and Dho, G. and {Di Giambattista}, F. and {Di Marco}, E. and Iacoangeli, F. and {Lima Júnior}, H. P. and Lopes, G. S. P. and Maccarrone, G. and Mano, R. D. P. and {Marcelo Gregorio}, R. R. and Marques, D. J. G. and Mazzitelli, G. and McLean, A. G. and Messina, A. and Monteiro, C. M. B. and Nobrega, R. A. and Pains, I. F. and Paoletti, E. and Passamonti, L. and Petrucci, F. and Piacentini, S. and Piccolo, D. and Pierluigi, D. and Pinci, D. and Prajapati, A. and Renga, F. and {J.d.C. Roque}, R. and Rosatelli, F. and Russo, A. and Saviano, G. and Spooner, N. J. C. and Tesauro, R. and Tomassini, S. and Torelli, S. and {dos Santos}, J. M. F.}
}

@article{SAULI1997531,
title = {GEM: A new concept for electron amplification in gas detectors},
journal = {Nuclear Instruments and Methods in Physics Research Section A: Accelerators, Spectrometers, Detectors and Associated Equipment},
volume = {386},
number = {2},
pages = {531-534},
year = {1997},
issn = {0168-9002},
doi = {https://doi.org/10.1016/S0168-9002(96)01172-2},
url = {https://www.sciencedirect.com/science/article/pii/S0168900296011722},
author = {F. Sauli}
}

@article{dho2024ely,
  title={Enhancing the light yield of {He}: {CF}$_4$ based gaseous detector},
  author={Dho, G and others},
  journal={The European Physical Journal C},
  volume={84},
  number={10},
  pages={1122},
  year={2024},
  publisher={Springer},
  doi={10.1140/epjc/s10052-024-13471-5}
}

@article{pains2023idbscan,
doi = {10.1088/1361-6501/acf402},
url = {https://dx.doi.org/10.1088/1361-6501/acf402},
year = {2023},
month = {sep},
publisher = {IOP Publishing},
volume = {34},
number = {12},
pages = {125024},
author = {Pains, I. F. and others},
title = {Directional {iDBSCAN} to detect cosmic-ray tracks for the {CYGNO} experiment},
journal = {Measurement Science and Technology}
}

@article{dimarco2020identification,
doi = {10.1088/1361-6501/abbd12},
url = {https://dx.doi.org/10.1088/1361-6501/abbd12},
year = {2020},
month = {dec},
publisher = {IOP Publishing},
volume = {32},
number = {2},
pages = {025902},
author = { Di Marco, E. and others},
title = {Identification of low energy nuclear recoils in a gas time projection chamber with optical readout},
journal = {Measurement Science and Technology}
}

@article{bertone2005particle,
title = {Particle dark matter: evidence, candidates and constraints},
journal = {Physics Reports},
volume = {405},
number = {5},
pages = {279-390},
year = {2005},
issn = {0370-1573},
doi = {10.1016/j.physrep.2004.08.031},
url = {https://www.sciencedirect.com/science/article/pii/S0370157304003515},
author = {Bertone, G. and Hooper, D. and Silk, J.}
}

@article{zurek2024,
   author = "Zurek, Kathryn M.",
   title = "Dark Matter Candidates of a Very Low Mass", 
   journal= "Annual Review of Nuclear and Particle Science",
   year = "2024",
   volume = "74",
   number = "Volume 74, 2024",
   pages = "287-319",
   doi = "https://doi.org/10.1146/annurev-nucl-101918-023542",
   url = "https://www.annualreviews.org/content/journals/10.1146/annurev-nucl-101918-023542",
   publisher = "Annual Reviews",
   issn = "1545-4134",
   type = "Journal Article",
  }

@article{mayet2016review,
title = {A review of the discovery reach of directional {Dark} {Matter} detection},
journal = {Physics Reports},
volume = {627},
pages = {1-49},
year = {2016},
note = {A review of the discovery reach of directional Dark Matter detection},
issn = {0370-1573},
doi = {10.1016/j.physrep.2016.02.007},
url = {https://www.sciencedirect.com/science/article/pii/S0370157316001022},
author = {F. Mayet and others}
}

@article{amaro2022cygno,
AUTHOR = {Amaro, F. D. and others},
TITLE = {The {CYGNO} {Experiment}},
JOURNAL = {Instruments},
VOLUME = {6},
YEAR = {2022},
NUMBER = {1},
ARTICLE-NUMBER = {6},
URL = {https://www.mdpi.com/2410-390X/6/1/6},
ISSN = {2410-390X},
DOI = {10.3390/instruments6010006}
}

@article{Morozov_2012,
	doi = {10.1088/1748-0221/7/02/p02008},
	url = {https://doi.org/10.1088/1748-0221/7/02/p02008},
	year = 2012,
	month = {feb},
	publisher = {{IOP} Publamountishing},
	volume = {7},
	number = {02},
	pages = {P02008--P02008},
	author = {A Morozov and L M S Margato and M M F R Fraga and L Pereira and F A F Fraga},
	title = {Secondary scintillation in CF$_4$: emission spectra and photon yields for {MSGC} and {GEM}},
	journal = {Journal of Instrumentation}
}

@incollection{IPCC_AR6_WGI_Chapter7_2021,
  author    = {{IPCC}},
  title     = {Chapter 7: The Earth’s Energy Budget, Climate Feedbacks and Climate Sensitivity},
  booktitle = {Climate Change 2021: The Physical Science Basis. Contribution of Working Group I to the Sixth Assessment Report of the Intergovernmental Panel on Climate Change},
  year      = {2021},
  publisher = {Cambridge University Press},
  note      = {See Supplementary Material, Table 7.SM.7 (Greenhouse gas lifetimes, radiative efficiencies, and climate metrics including GWP).},
  url       = {https://www.ipcc.ch/report/ar6/wg1/downloads/report/IPCC_AR6_WGI_Chapter_07_Supplementary_Material.pdf}
}

@misc{hamamatsu_r1635,
  author       = {{Hamamatsu Photonics}},
  title        = {{R1635 Photomultiplier Tube}},
  howpublished = {\url{https://www.hamamatsu.com/us/en/product/optical-sensors/pmt/pmt_tube-alone/head-on-type/R1635.html}},
  note         = {Accessed: 2025-02-04}
}

@misc{hamamatsu_orca_quest2,
  author       = {{Hamamatsu Photonics}},
  title        = {{ORCA-Quest 2 qCMOS Camera C15550-22UP}},
  howpublished = {\url{https://www.hamamatsu.com/us/en/product/cameras/qcmos-cameras/C15550-22UP.html}},
  note         = {Accessed: 2025-02-04}
}

@misc{ehd_f085_lens,
  author       = {{EHD imaging GmbH}},
  title        = {{EHD-F085 Lens}},
  howpublished = {\url{https://www.ehd.de/wp-content/uploads/2024/05/EHD-F085-Lenses.pdf}},
  note         = {Accessed: 2025-02-04}
}

@article{methane,
      title={He-CF4-CH4 ternary mixtures as target gas for the CYGNO directional dark matter experiment},
      author={F. D. Amaro and R. Antonietti and E. Baracchini and L. Benussi and C. Capoccia and M. Caponero and L. G. M. de Carvalho and G. Cavoto and I. A. Costa and A. Croce and M. D'Astolfo and G. D'Imperio and G. Dho and E. Di Marco and J. M. F. dos Santos and D. Fiorina and F. Iacoangeli and Z. Islam and H. P. Lima Jr and G. Maccarrone and R. D. P. Mano and D. J. G. Marques and G. Mazzitelli and P. Meloni and A. Messina and C. M. B. Monteiro and R. A. Nobrega and I. F. Pains and E. Paoletti and F. Petrucci and S. Piacentini and D. Pierluigi and D. Pinci and F. Renga and R. J. da C. Roque and A. Russo and G. Saviano and P. A. O. C. Silva and R. C. Sousa and N. J. C. Spooner and R. Tesauro and S. Tomassini and D. Tozzi},
      year={2026},
      journal={arXiv preprint arXiv:2606.01512},
      eprint={2606.01512},
      archivePrefix={arXiv},
      primaryClass={physics.ins-det},
      url={https://arxiv.org/abs/2606.01512},
}

@article{Billard2022APPEC,
      author        = "Billard, Julien and Boulay, Mark and Cebrián, Susana and
                       Covi, Laura and Fiorillo, Giuliana and Green, Anne and
                       Kopp, Joachim and Majorovits, Béla and Palladino, Kimberly
                       and Petricca, Federica and Roszkowski, Leszek and Schumann,
                       Marc",
      title         = "{Direct detection of dark matter—APPEC committee report}",
      archivePrefix = "arXiv",
      eprint        = "2104.07634",
      journal       = "Rep. Prog. Phys.",
      volume        = "85",
      number        = "5",
      pages         = "056201",
      year          = "2022",
      url           = "https://cds.cern.ch/record/2764484",
      note          = "106 pages",
      doi           = "10.1088/1361-6633/ac5754",
}

@article{Trex,
author={Castel, J.
and Cebri{\'a}n, S.
and Coarasa, I.
and Dafni, T.
and Gal{\'a}n, J.
and Iguaz, F. J.
and Irastorza, I. G.
and Luz{\'o}n, G.
and Mirallas, H.
and Ortiz de Sol{\'o}rzano, A.
and Ruiz-Ch{\'o}liz, E.},
title={Background assessment for the TREX dark matter experiment},
journal={The European Physical Journal C},
year={2019},
month={Sep},
day={21},
volume={79},
number={9},
pages={782},
issn={1434-6052},
doi={10.1140/epjc/s10052-019-7282-6},
url={https://doi.org/10.1140/epjc/s10052-019-7282-6}
}

@article{BENZIGER2007509,
title = {The nylon scintillator containment vessels for the Borexino solar neutrino experiment},
journal = {Nuclear Instruments and Methods in Physics Research Section A: Accelerators, Spectrometers, Detectors and Associated Equipment},
volume = {582},
number = {2},
pages = {509-534},
year = {2007},
issn = {0168-9002},
doi = {https://doi.org/10.1016/j.nima.2007.08.176},
url = {https://www.sciencedirect.com/science/article/pii/S0168900207018712},
author = {J. Benziger and L. Cadonati and F. Calaprice and E. {de Haas} and R. Fernholz and R. Ford and C. Galbiati and A. Goretti and E. Harding and A. Ianni and An. Ianni and S. Kidner and M. Leung and F. Loeser and K. McCarty and A. Nelson and R. Parsells and A. Pocar and T. Shutt and A. Sonnenschein and R.B. Vogelaar}
}

@article{Ghrear_2021,
doi = {10.1088/1475-7516/2021/10/005},
url = {https://doi.org/10.1088/1475-7516/2021/10/005},
year = {2021},
month = {oct},
publisher = {IOP Publishing},
volume = {2021},
number = {10},
pages = {005},
author = {Ghrear, M. and Vahsen, S.E. and Deaconu, C.},
title = {Observables for recoil identification in high-definition Gas Time Projection Chambers},
journal = {Journal of Cosmology and Astroparticle Physics}
}

@article{Beaufort_2024,
doi = {10.1088/1748-0221/19/05/P05052},
url = {https://doi.org/10.1088/1748-0221/19/05/P05052},
year = {2024},
month = {may},
publisher = {IOP Publishing},
volume = {19},
number = {05},
pages = {P05052},
author = {Beaufort, C. and Guillaudin, O. and Santos, D. and Sauzet, N. and Mobio, E. and Babut, R. and Tao, C.},
title = {Directional detection of keV proton and carbon recoils with MIMAC},
journal = {Journal of Instrumentation}
}

@article{NAMAI2025170145,
title = {Development of a low-background micro-pixel chamber for directional dark matter searches},
journal = {Nuclear Instruments and Methods in Physics Research Section A: Accelerators, Spectrometers, Detectors and Associated Equipment},
volume = {1072},
pages = {170145},
year = {2025},
issn = {0168-9002},
doi = {https://doi.org/10.1016/j.nima.2024.170145},
url = {https://www.sciencedirect.com/science/article/pii/S0168900224010714},
author = {Ryota Namai and Satoshi Higashino and Hirohisa Ishiura and Tomonori Ikeda and Mizuno Ofuji and Ayaka Nakayama and Ryo Kubota and Kiseki D. Nakamura and Hiroshi Ito and Koichi Ichimura and Ko Abe and Kazuyoshi Kobayashi and Atsushi Takada and Kentaro Miuchi}
}

\end{document}